# Single-step Chemical Vapor Deposition of Methyl Ammonium Lead Halide Perovskite for p-i-n Solar Cells

*Dmitry S. Muratov*[*,1], *Lev Luchnikov*[1], *Danila Saranin*[1], *Artur Ishteev*[1,3], *Vladislav Kurichenko*[2], *Evgeniy Kolesnikov*[1], *Denis V. Kuznetsov*[2] *and Aldo Di Carlo*[4]

[1]L.A.S.E. – Laboratory for Advanced Solar Energy, National University of Science and Technology "MISiS", Leninskiy prospect 6, Moscow, 119049, Russia

[2]Department of Functional NanoSystems and High-Temperature Materials, National University of Science and Technology "MISiS", Leninskiy prospect 4, Moscow, 119049, Russia

[3]N.N. Semenov Federal Research Center for Chemical Physics, Russian Academy of Sciences, Kosygina st. 4, Moscow, 119991, Russia

[4]CHOSE – Centre for Hybrid and Organic Solar Energy, Department of Electronic Engineering, University of Rome Tor Vergata, via del Politecnico 1, Rome, 00133, Italy

*Corresponding Author E-mail Address: muratov@misis.ru

**Abstract**

Metal halide perovskite solar cells being one of the fastest emerging technologies for renewable energy still has to become more industry friendly in a way that will allow using it for thin film modules or tandems with conventional silicon devices. The simplest way to achieve this is to use chemical vapor deposition (CVD) technique for tandem production. In this work, we show a method for a single step production of $MAPbI_3$ films in a simple two-zone CVD reactor from lead diacetate and methyl-ammonium iodide powders. Obtained films show highly ordered cubic $MAPbI_3$ phase with a thickness of 400-500 um, good photoluminescence response and absorption band edge similar to spin-coated film. We used those films to produce p-i-n solar cells with $ITO/NiO/MAPbI_3/C60/Cu$ structure. The best cell showed negligible 0.23 % PCE right after the manufacturing but significantly improved to 5.5 % PCE after 8 hours of storage in the dark. The main limiting factor affecting the efficiency is low current density, which we attribute to non-optimized growth conditions; however, our approach is a first step to a single step CVD deposition of $MAPbI_3$ on any type of substrates including texturized silicon subcells for better overall efficiency.

**Introduction**

Metal-organic halide perovskites (HP) have shown an impressive performance leap in thin-film solar cell technology over the last 10 years. The most famous member of the HP family is the methyl-ammonium lead iodide $(MAPbI_3)$[1], which has been studied, characterized, and widely used for fabrication of perovskite solar cells with efficiency even exceeding the 25%[2]. More recently the use of multication perovskite such as CsMAFAPbI$_3$ or CsFAPbI$_3$[3,4] with reduced bandgap permitted to increase cell efficiency and improve stability[5]. Nevertheless, single cation MA-based perovskites are still of the interest and their use in different PV applications has been considered[6]. One of the main successes of MAPbI$_3$ are its simple deposition process based of solution-process coating techniques such as spin-coating[7] slot-die printing[8], screen-printing[9] and blade-coating techniques[10]. At the same time, physical deposition techniques such as thermal sublimation has been proven to be compatible with halide perovskite deposition especially when conformable coating is necessary an in the case of tandem perovskite/silicon where PSC is deposited on a texturized silicon surface[11]. Among different physical deposition technique, the chemical vapor deposition (CVD) method is one of the most suited ad industry level and well recognized for conventional PV technology[12,13] Most of the published research on MAPbI$_3$ CVD used either spin-coated PbI$_2$ films[14,15] as a required preliminary step for photoactive layer production or involve thermal evaporation of PbI$_2$ in high-vacuum conditions[16]. These routes produce very good quality films but getting similar results in a single step reaction will yield a more industry friendly method ready to use in already available equipment or for silicon-perovskite tandem production. The only paper describing full CVD produced solar cell in single-step manner used an n-i-p structure with FTO/TiO$_2$ layer being in a hot zone of the furnace for perovskite deposition[17] but the extension to the inverted structure where the HP is deposited on a hole transporting material (HTM) has not been demonstrated. Considering the importance, the p-i-n structure is gaining in the recent time, especially for the tandem perovskite/silicon[18] we focus our attention in developing a single-step CVD process suitable for the p-i-n configuration. In doing this, we also take care to the thermal budged of the process considering that the HTM in p-i-n structure is typically an organic material such as PEDOT:PSS, PTAA or a Self-Assembling organic Monolayer (SAM).

**Experimental**

We used chemically pure 99.99% lead iodide (PbI$_2$) (Lanhit), lead diacetate trihydrate (Pb(CH$_3$COO)$_2$3H$_2$O) 98% (Lenreactiv) and methylammonium iodide (CH$_3$NH$_3$I) (Great Cell). We used pure argon (99.9995 %) as a carrier gas with mass flow controllers in two types of quartz tubes 25 mm and 50 mm in diameter. In a typical synthesis procedure, the precursor materials are put into separate quartz boats inside a bigger quartz tube reactor with a single heated area by the tube furnace (Carbolite EST 1200) for the 25 mm tubes (**fig. 1a**). In case of separate evaporation and deposition zones (50 mm tube) we used homemade chromel heater over evaporation zone and the tube furnace for the deposition zone. Solar cells with p-i-n configuration were made on patterned glass/ITO/NiO substrates, where NiO layer is spin-coated from NiCl solution as described elsewhere[19]. Photoactive layer was then covered by thermally evaporated C60 at 10$^{-6}$ Torr vacuum level. Bathocuproine interlayer was spin coated at 4000 RPMs (30s) and annealed at 50 °C (1 min) for reference devices. Copper cathode was deposited also with thermal evaporation method trough shadow mask to form 0.15 cm$^2$ active area for each cell.

To analyze the structure and morphology of obtained films we used scanning electron microscopy (SEM) images obtained with Tescan Vega 3. Molecular structure was analyzed with Fourier-transform infrared (FTIR) spectrometer (Thermo Nicolet) in attenuated total reflectance mode with ZnSe crystal. Crystal structure and phase composition was analyzed by X-Ray diffraction method with XRD patterns measured on Difrey instrument with Cr Kα radiation (0.229 nm). Absorbance and photoluminescence spectra of MAPbI$_3$ films were measured using Shimadzu UV-Viz 1280 and Carry Eclipse spectrometer with 550 nm excitation wavelength. The photovoltaic (PV) parameters for fabricated PSCs (open-circuit voltage - Voc, short circuit current density - Jsc, fill factor – FF, and power conversation efficiency – PCE) were calculated from the IV characteristics measured under standard conditions of 1.5 AM G light spectrum. We used Xenon arc-lamp based solar simulator (ABET 3000) with AAA grade of conformity to the reference Air Mass 1.5 Global spectrum of the terrestrial solar light (1.5 AM G standard - (ASTM) G-173). The standard power density of the incoming light (100 mW/cm$^2$) was calibrated with certified Si- reference solar cell. The scan sweeps and the MPPT tuning measurements were performed with Keithley 2400 SMU.

**Results and Discussion**

We compared two types of precursors for lead sources, namely lead iodide (PbI$_2$) and lead diacetate (PbAc$_2$) with CH$_3$NH$_3$I for iodide and methyl-amine groups respectively. Lead iodide as a precursor is widely discussed in literature[19] being a relatively stable reagent leaving no additional components after the formation of MAPbI$_3$ structure. For CVD approach, however, PbI$_2$ is not very convenient due to very different melting points of PbI$_2$ and MAI (402 °C and 270 °C, respectively)[20]. However, beside PbI$_2$, formation of MAPbI$_3$ perovskite with cubic structure can be obtained from other precursors such as lead diacetate (PbAc$_2$)[21], lead oxides[22] or pure metallic lead[23]. To obtain a MAPbI$_3$ film with a single zone CVD furnace we needed to use a precursor with similar melting and evaporation temperatures to MAI and this condition is well met by PbAc$_2$. We should point out that even chemically pure lead diacetate if stored in ambient conditions will eventually become lead diacetate trihydrate (Pb(CH$_3$COO)$_2$·3H$_2$O) which melts at 75 °C in its crystalline water and then decomposes at 200 °C . This process does not allow forming a stable MAPbI$_3$ structure and for this reason, PbAc$_2$ must be freeze dried before use to remove all the crystalline water [24].

Several experiments have been performed to compare the CVD growth of MAPbI$_3$ obtained with two types of precursors. If the CVD is performed in argon flow but without vacuum conditions, we either obtained unstable brown film which became opaque white when taken out from the furnace or MAI reacted with PbAc$_2$ in the crucible producing lead oxides on the tube and on the substrates. To avoid this problem a vacuum conditions (10$^{-1}$ mbar) were introduced combined with fixed argon flow. For the experiments with 25 mm tubes we used stable 75 sccm argon flow.

We compared both PbI$_2$ and PbAc$_2$ precursors with the same synthesis parameters such as substrate positions and gas flow to analyze the film quality, its composition, morphology and optical properties. The temperature in the hot zone of the furnace for PbAc$_2$ was 200 °C and for the PbI$_2$ 350 °C due to different melting points. The cold zone of the furnace has the temperature 50°C lower compared to the hot zone (**fig. 1a**). For PbAc$_2$ it drops to 150 °C and for PbI$_2$ to 200 °C resulting in gradual change of film composition with the length of the tube reactor. Most of the experiments were done with two substrates (12.5 mm x 25 mm) placed one after another to observe the formation of all the phases for further analysis. The first substrate from the precursor crucible along the gas flow has two different color zones: the first bright yellow film is mainly related to PbI$_2$ as measured by XRD while the second zone has a dark brown color that is MAPbI$_3$. The next substrate in most of the experiments is also of dark brown color but has different morphology and it has high excess of iodine. Scanning electron microscopy (SEM) images from the dark part of the first substrates for PbAc$_2$ and PbI$_2$ precursors are shown in **Fig. 1b** and **c**, respectively. The morphology of the film obtained from PbAc$_2$ has apparent cubic structure and the film itself does not have any cracks or pores. Mean grain size slightly varies from 0.45 to 0.55 μm and the film thickness is 400-500 nm as measured by stylus profilometry. The dark films obtained from PbI$_2$ has larger crystals and very rough surface. The grain size varies from 15 to 25 μm, but at higher magnification (right inlet of **Fig. 1c**) we noticed small hexagonal flakes which resemble the structure of PbI$_2$. To analyze the crystal structure of the film for the different zones and for the two precursor we performed X-Ray diffraction (XRD) measurements. XRD patterns are shown in **Fig. 2** together with the images of substrates.

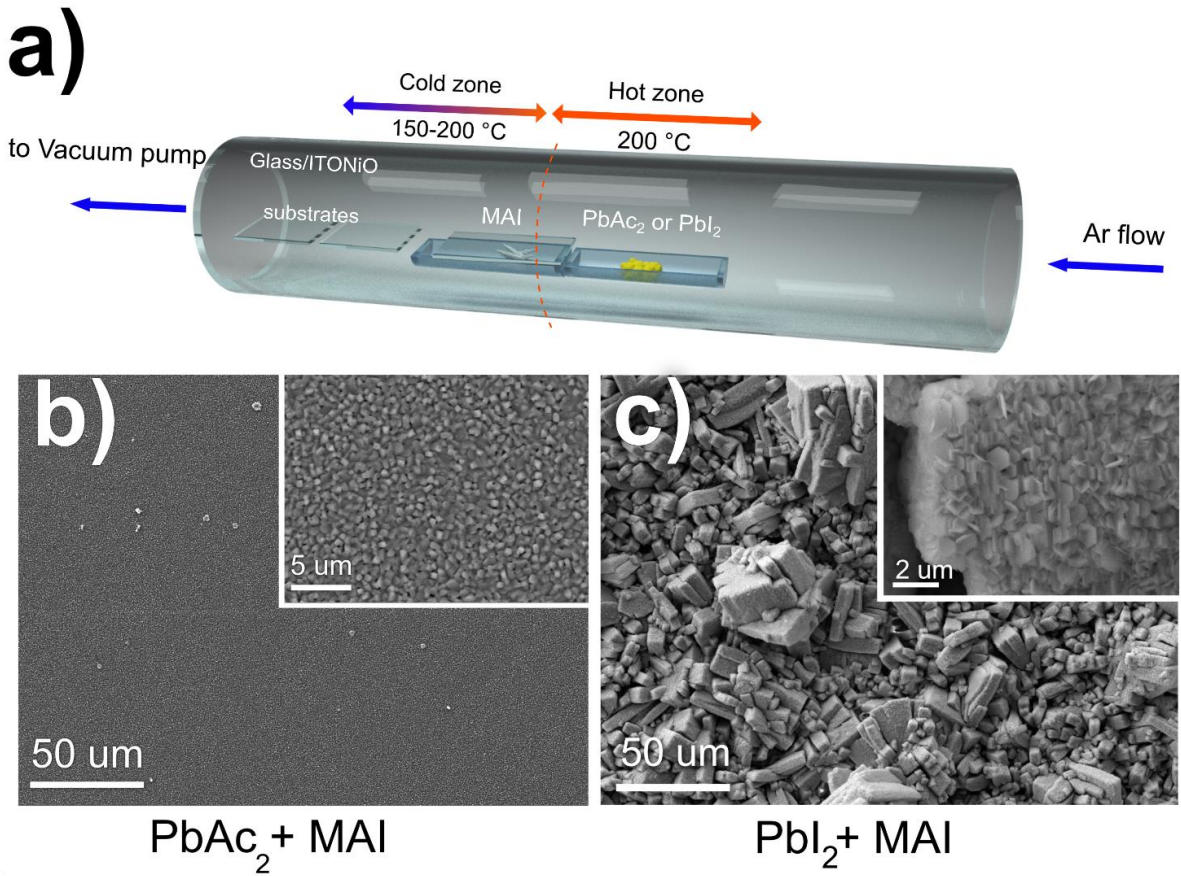

**Figure 1.** CVD process scheme (a) showing two zones of the furnace and precursor placements for synthesis with PbI$_2$ we placed it in hot zone and MAI on the cooler side, for the synthesis with PbAc$_2$ we swapped the precursors to place MAI in the hot zone and PbAc$_2$ on the cooler side SEM images of the films obtained from lead acetate (b) and lead iodide (c) show different morphology and grain size.

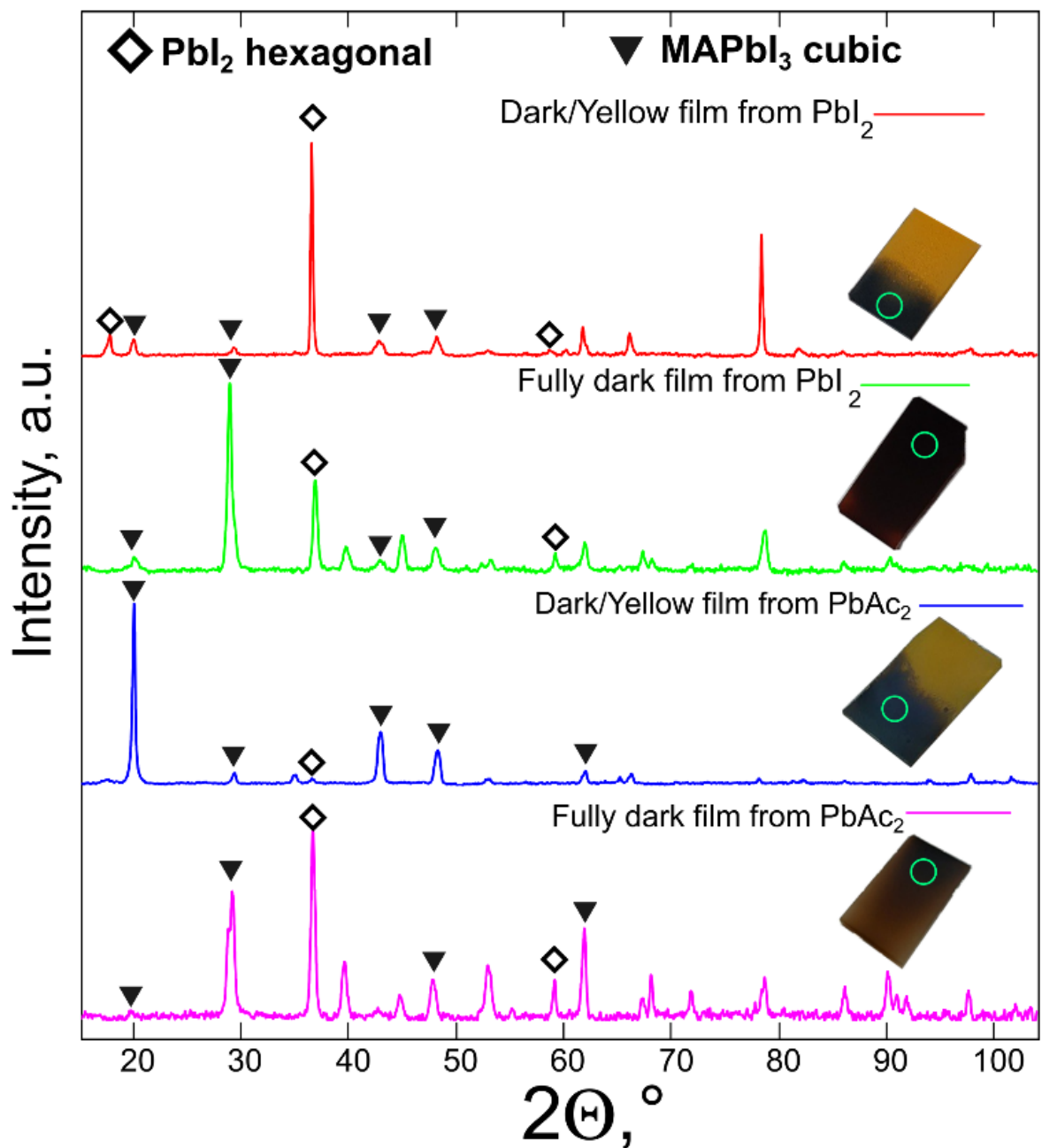

**Figure 2.** XRD patterns of obtained films on glass from two types of precursors (PbAc$_2$ and PbI$_2$). The dark/yellow films were placed closer to the hot zone in the furnace and fully dark films are formed further away from it. The diamond marks hexagonal PbI$_2$ peaks and triangle marks the peaks of cubic MAPbI$_3$ phase. The spot from which the pattern was measured is marked with a green circle on all the substrates.

Here the green circles mark the place of the measurement. The two-colored (dark/yellow) film obtained from PbI$_2$ precursor has both PbI$_2$ and MAPbI$_3$ cubic phases, even though the peaks of PbI$_2$ have higher intensity. Fully dark film obtained with the PbI$_2$ precursor have higher content of MAPbI$_3$ but PbI$_2$ peaks are still present. The most interesting pattern was obtained from the dark part of two-colored (dark/yellow) film when PbAc$_2$ is used as precursor. As shown in **Fig. 2** (blue line) the peaks of cubic MAPbI$_3$ phase have very high intensity with almost no additional phases present. That film also has the best morphology and a smooth surface according to SEM data shown in **Fig. 1b**. Fully dark film obtained from PbAc$_2$ is similar to the dark film from PbI$_2$ having both PbI$_2$ and MAPbI$_3$ phases. We noticed that this dark film from PbAc$_2$ can change the color to opaque white when exposed to ambient conditions, even though the dark state could be revived by annealing at 100 °C indicating the occurrence of a phase transition. This effect can be explained by the formation of a hydrated MAPbI$_3$·H$_2$O phase in ambient conditions. The rate of hydrated phase formation depends on the amount of unreacted MAI and the degree of conversion of MAPbI$_3$ to cubic perovskite[25].

To confirm the molecular structure of the films, we performed Fourier-transformed infrared (FTIR) spectroscopy (**Fig. 3a**). We compared the spectra of CVD produced films (dark part of dark/yellow substrates) for both PbI$_2$ and PbAc$_2$ precursors as well as MAPbI$_3$ powder obtained by ball milling. The spectra have all the bands for MAPbI$_3$ described by Abdelmageed et. al.[26] On all the spectra the highest intensity band at 906 cm$^{-1}$ is attributed to CH$_3$NH$_3$ rocking vibrations, the band at 1460 cm$^{-1}$ is attributed to NH$_3$ bending vibrations, all the bands in the region from 3040 to 3085 cm$^{-1}$ are attributed to N-H bond stretching. The bands from C-H bond stretching are present but have very low intensity. The main differences of the spectra from the films obtained from PbI$_2$ and PbAc$_2$ bands are at 1400 cm$^{-1}$ and 1250 cm$^{-1}$ when the strongest bands at those frequencies are for the films from PbI$_2$ precursor. The spectrum from PbAc$_2$ showed the lowest intensity due to low film thickness and was calculated as the difference between the film on glass and pristine glass spectra.

We also analyzed the absorption of the dark films by UV-Vis spectroscopy (**Fig. 3b**). The film obtained from PbAc$_2$ shows strong absorption band edge at 780 nm, which is typical for MAPbI$_3$, on the other hand the film obtained from PbI$_2$ have the absorption band edge shifted to almost 800 nm. According to X. M. Zhang et al.[27] red shift could be due to iodine vacancies and probably partial phase transition from cubic/orthorombic to trigonal or monoclinic phases We concluded that conditions needed for the formation of a good film from PbAc$_2$ with cubic MAPbI$_3$ are well met only at a very narrow area of the furnace.

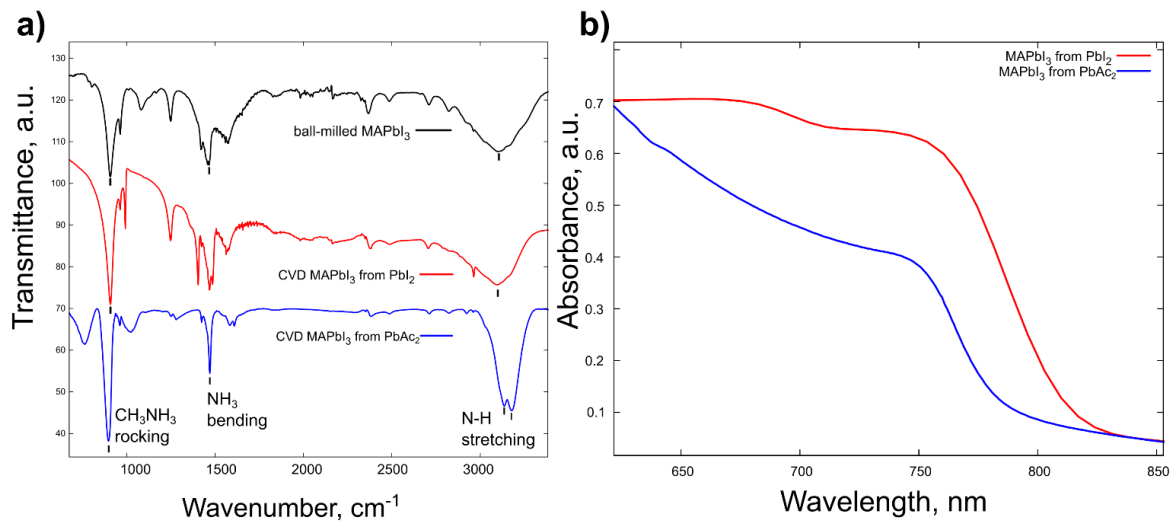

**Figure 3.** FTIR spectra (a) of ball-milled MAPbI$_3$ and CVD MAPbI$_3$ from PbI$_2$ and PbAc$_2$ precursors and absorbance spectra for the films from PbI$_2$ and PbAc$_2$ showing a shift of absorption band edge from 780 to 800 nm (b)

Owing to the information gathered in these preliminary experiments, we re-design the CVD deposition chamber considering i) substrates placed perpendicular to the gas flow, ii) a larger furnace diameter (50 mm) to increase the properly place the substrate and iii) a second heated zone and separate carrier gas flow rate for the precise control over film deposition conditions. For this scheme, we used 75 sccm flow for $PbAc_2$ source and 15 sccm for MAI source. The scheme of improved process as well as film and device photographs with PL, UV-Viz data, XRD pattern and SEM images are shown in **Fig. 4**. The optimized deposition scheme increased a number of pixels per single substrate and raised a number of substrates in the single process. Slight shadowing effect for the substrates placed further from the sources does occur, but all the substrates are fully covered with dark $MAPbI_3$ phase, which is expected for a CVD process[28]. The films produced in such process have absorbance spectra similar to what was achieved in previous scheme (**fig.4d**) and show a PL peak at 780 nm, which is comparable to what is reported for conventional spin-coated films. Mean thickness of the films is 450±50 nm and should be optimal for device preparation. An XRD pattern of the films have all characteristic peaks of cubic $MAPbI_3$ (COD: 7225287) and no peaks of $PbI_2$ phase, that could affect the device performance and increase degradation rate as well. SEM images show lower grain size and smoother looking surface (**Fig. 4f**) if compared to the films produced in the first iteration of our CVD scheme.

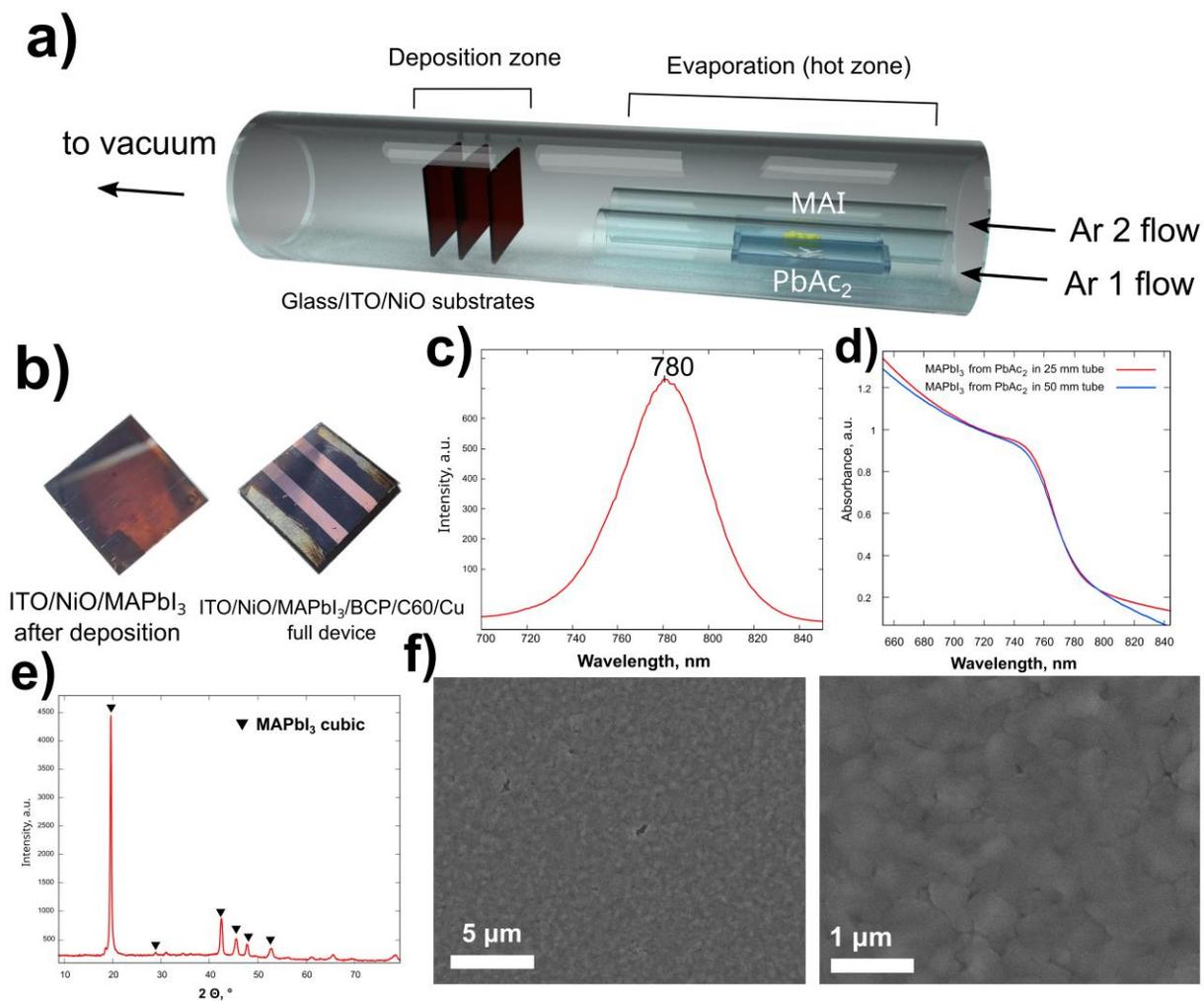

**Figure 4**. (a) Optimized CVD process scheme, (b) photos of MAPbI$_3$ film and same film after device assembly, (c) PL spectrum of the film after deposition, (d) Absorbance spectra of MAPbI$_3$ produced with the first scheme (red) and with optimized scheme (blue), (e) XRD pattern of CVD grown MAPbI$_3$ film, (f) SEM images of the films produced on bigger substrates in optimized process

After assessment of structural and optical properties of MAPbI$_3$ thin-films fabricated with CVD approach, we manufactured solar cells with p-i-n architecture. The inverted device structure was realized using the following stack: glass-ITO (pixelated substrate)/NiO$_x$ (hole transporting layer, 20 nm)/perovskite absorber (CVD processing)/$C_{60}$(electron transport layer, 40 nm)/BCP(hole blocking interlayer, 8 nm)/Copper (cathode, 100 nm). The measurements of the output IV performance were obtained by using the AAA class Solar simulator (Xe Source, 1.5 AM G spectrum calibrated with certified Si cell). Fabricated perovskite solar cells demonstrated interesting performance behavior with strong correlation to the time of light-soaking (LS). As fabricated, devices showed poor performance with open circuit voltage ($V_{oc}$) = 0.85 V, short circuit current density ($J_{sc}$) = 0.6 mA/$cm^2$, fill factor (FF) = 45 % and power conversion efficiency (PCE) = 0.23 %. By exposing the PSC to continuous light-soaking (light power irradiance density – 100 mW/$cm^2$ and heating up to 63±1.5 ˚C) improved the PCE of PCSs mainly due to an increase of $J_{sc}$. As it is showed in **fig.5**, the $J_{sc}$ value increased up to 1.39 mA/$cm^2$ after 10 minutes of LS and up to 4.71 mA/$cm^2$ after 60 minutes. The storage in ambient conditions in the dark also improves the performance. The $J_{sc}$ values of the PSC after dark storage (8 hours) incremented up to 10.8 mA/$cm^2$ and showed 5.54 % PCE. This improvement we address to either slow recrystallization or internal stress relaxation of the photoactive layer. After the encapsulation, we observed slightly lower performance, but $V_{oc}$ increased from 0.87 to 0.95 V, which is similar to spin-coated devices of the same configuration. We analyzed the slope of JV curve near $V_{oc}$ point and extracted the values of the series resistance ($R_s$). The transport properties of "as fabricated" PSC were affected with huge $R_s$ value of $4.27 \cdot 10^2$ Ohm·$cm^2$. Under LS conditions the series resistance reduced up to $6 \cdot 10^1$ Ohm·$cm^2$ after 60 minutes. Interestingly, that after dark storage of the devices, series resistance continued the reduction to $2.2 \cdot 10^1$ Ohm $cm^2$, while encapsulation processing the value of $R_s$ returned to the $6 \cdot 10^1$ Ohm·$cm^2$.

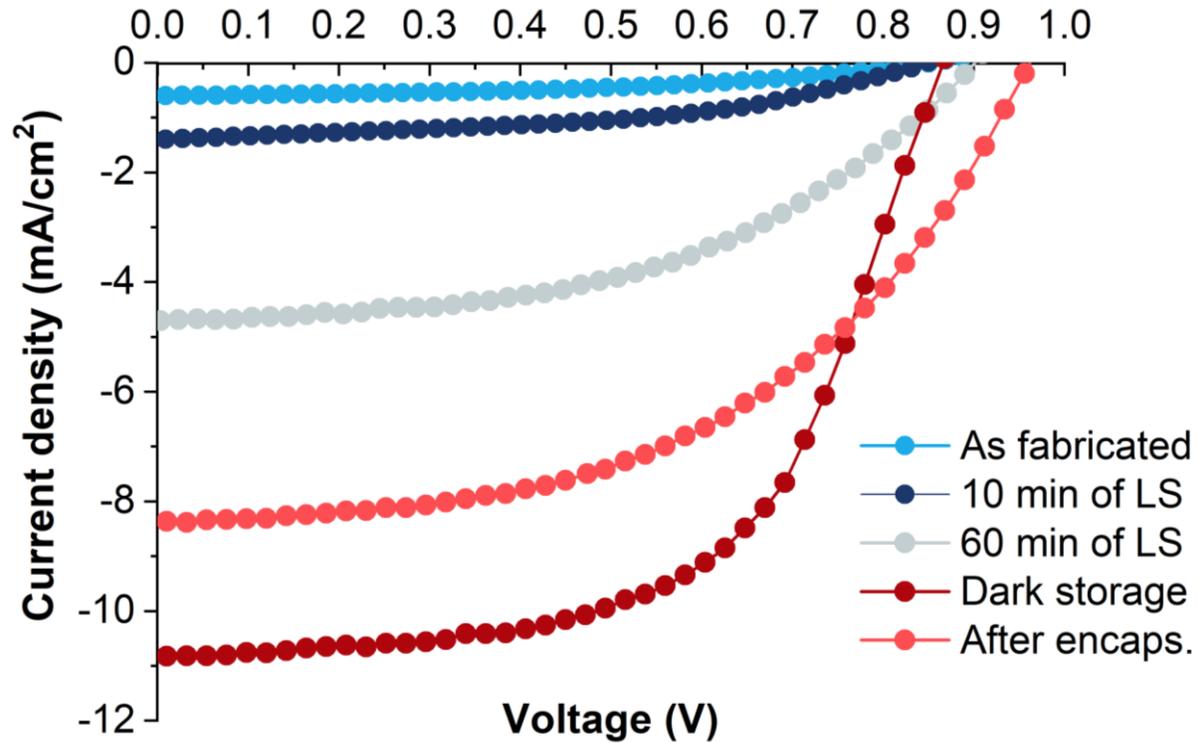

**Figure 5.** The best JV-curves of devices with CVD produced MAPbI$_3$ photoactive layer

We observed three main changes of the J-V characteristics induced by the post-fabrication treatments: a) an improvement of the photocurrent during LS; b) the boost of photocurrent after dark storage, and c) the increase of $V_{oc}$ after encapsulation with ultraviolet exposure. The stability of the device performance under continuous light soaking is an important topic for PV and specific features exists for halide perovskite solar cells [29,30]. The reports on LS highlight the critical role of the light-activated trap-states[31,32], self-healing defects[33,34], ion migration [35] and structural meta-stability[36]. Considering methyl-ammonium based perovskites, it is necessary to take into account the relatively low temperature of the photoactive phase transition (55°C)[37], which is close to the conditions of the solar cell operation under light illumination and encapsulation. As reported for p-i-n structured PSCs fabricated with solution processing[38], the steady state conditions, when device reaches peak power output, requires short period of LS. This process accompanied with filling of the light activated traps that are represented by shallow and deep-state defects – iodine vacancies[39], iodine interstitials[40], and antisites of iodine with organic cation[40]. In our case, we observed that device require at least 60 minutes to compensate the negative impact of the intrinsic point defects, and probably, finish the phase transition of the absorber film. The improvement of the device performance after dark storage shows, that such condition induces the additional recovery of perovskite layer. As reported by *Cheng et al.*[38], the dark storage conditions allows the dissociation of the ionic defects accumulated at the device's interfaces and restoration of the vacancies in the crystal lattice of perovskite. The encapsulation resulted in decrease of the performance originated from the decreased $J_{sc}$ and incremented $R_s$. We assume that continuous UV-exposure accelerated the formation of the mid-gap light-activated trap and accumulation at the interfaces. The observed changes in the IV performance of the CVD processed PSCs clearly demonstrates that semiconductor properties of the absorber films are meta-stable and sensitive to the light treatment and room temperature storage. The stabilization of the phase transition and control of the defect formation for the CVD perovskite thin-films obviously requires advanced methods for the monitoring of the fabrication process.

**Conclusions**
In this work, we describe a simple single-step CVD process for the deposition of $MAPbI_3$ films from solid state precursors both on top of glass and on ITO/NiO substrates. The best films were obtained from lead acetate and methyl-ammonium iodide due to similar melting points. The process we developed produce a smooth film with mean grain size of 0.5 μm and thickness of 450-500 nm. Most of the surface of the film considered cubic phase of $MAPbI_3$ and can be stable in ambient conditions for several days. The best overall performance of p-i-n MAPbI3 solar cells of 5.54 % required an implementation of separate control over carrier gas above each of the precursor powders as well as two-zone heating regime. We also report the formation of metastable dark perovskite-like film at lower deposition temperature. This film undergoes a transition to opaque white film when stored in air, has residual $PbI_2$ and an excess of iodine according to EDX. When heated to 100 °C this phase quickly returns to dark state. We assume that this phase transition is probably due to transformation from α` to ε phase but the temperature is lower due to higher MAI content in the film

**Corresponding Author**

**Dr. Dmitry S. Muratov: muratov@misis.ru**

**Prof. Aldo DI Carlo aldo.dicarlo@uniforma2.it**

**Author Contributions**

**Dmitry S. Muratov**: Conceptualization, Methodology, Formal analysis, Data curation, Investigation, Visualization, Writing- Original draft preparation. **Lev Luchnikov:** Methodology, Formal Analysis, Investigation, Visualization**. Artur Ishteeev:** Conceptualization, Resources. **Danila Saranin:** Validation, Resources, Data curation, Writing – Review & Editing. **Vladislav Kurichenko:** Methodology, Investigation, Formal analysis. **Evgeny Kolesnikov:** Investigation, Formal analysis, Visualization. **Denis V. Kuznetsov:** Conceptualization, Supervision, Project administration**. Aldo Di Carlo:** Conceptualization, Writing – Original Draft, Review & Editing, Supervision, Project administration, Funding acquisition.

ACKNOWLEDGMENT

The authors gratefully acknowledge the financial support of the Russian Science Foundation (RSF) under grant number 21-19-00853